\documentclass[aps,prl,twocolumn,superscriptaddress,floatfix,showpacs]{revtex4}
\usepackage[dvipdfmx]{graphicx}
\usepackage[dvipdfmx]{color}
\usepackage{amsmath}
\usepackage{amssymb}
\usepackage{dcolumn}
\usepackage{bm}
\usepackage{latexsym} 
\usepackage{setspace}
\usepackage{graphicx}
\usepackage{color}
\begin{document}
\title{Detection of induced paramagnetic moments in Pt on Y$_3$Fe$_5$O$_{12}$ \\
via x-ray magnetic circular dichroism}
\author{Takashi Kikkawa}
\email{t.kikkawa@imr.tohoku.ac.jp}
\affiliation{Institute for Materials Research, Tohoku University, Sendai 980-8577, Japan}
\affiliation{WPI Advanced Institute for Materials Research, Tohoku University, Sendai 980-8577, Japan}
\author{Motohiro Suzuki}
\affiliation{Japan Synchrotron Radiation Research Institute (JASRI), Sayo, Hyogo 679-5198, Japan}
\author{Jun Okabayashi}
\affiliation{Research Center for Spectrochemistry, The University of Tokyo, Bunkyo, Tokyo 113-0033, Japan}
\author{Ken-ichi Uchida}
\affiliation{Institute for Materials Research, Tohoku University, Sendai 980-8577, Japan}
\affiliation{National Institute for Materials Science, Tsukuba 305-0047, Japan}
\affiliation{PRESTO, Japan Science and Technology Agency, Saitama 332-0012, Japan}
\affiliation{Center for Spintronics Research Network, Tohoku University, Sendai 980-8577, Japan}
\author{Daisuke Kikuchi}
\email[Present address: School of Materials Science, Japan Advanced Institute of Science and Technology, Ishikawa 923-1292, Japan.]{}
\affiliation{Institute for Materials Research, Tohoku University, Sendai 980-8577, Japan}
\affiliation{WPI Advanced Institute for Materials Research, Tohoku University, Sendai 980-8577, Japan}
\author{Zhiyong Qiu}
\affiliation{WPI Advanced Institute for Materials Research, Tohoku University, Sendai 980-8577, Japan}
\author{Eiji Saitoh}
\affiliation{Institute for Materials Research, Tohoku University, Sendai 980-8577, Japan}
\affiliation{WPI Advanced Institute for Materials Research, Tohoku University, Sendai 980-8577, Japan}
\affiliation{Center for Spintronics Research Network, Tohoku University, Sendai 980-8577, Japan}
\affiliation{Advanced Science Research Center, Japan Atomic Energy Agency, Tokai 319-1195, Japan}
\date{\today}
\begin{abstract}
Magnetic moments in an ultra-thin Pt film on a ferrimagnetic insulator Y$_3$Fe$_5$O$_{12}$ (YIG) have been investigated at high magnetic fields and low temperatures by means of X-ray magnetic circular dichroism (XMCD). We observed an XMCD signal due to the magnetic moments in a Pt film at the Pt $L_{3}$- and $L_{2}$-edges. By means of the element-specific magnetometry, we found that the XMCD signal at the Pt $L_{3}$-edge gradually increases with increasing  the magnetic field even when the field is much greater than the saturation field of YIG. Importantly, the observed XMCD intensity was found to be much greater than the intensity expected from the Pauli paramagnetism of Pt when the Pt film is attached to YIG. These results imply the emergence of induced paramagnetic moments in Pt on YIG and explain the characteristics of the unconventional Hall effect in Pt/YIG systems.
\end{abstract}
%75.70.-i: Magnetic properties of thin films, surfaces, and interfaces
%72.25.-b: Spin polarized transport
%75.47.-m: Magnetotransport phenomena; materials for magnetotransport
%85.75.-d: Magnetoelectronics; spintronics: devices exploiting spin polarized transport or integrated magnetic fields 
%
\maketitle
%
%------------main-text----------------------------------
%
%%%%%%%%%%%%%%%%%%%%%%%%%%%%%%%%%%%%%%%%%%%%%%%%%%%%
\section{I.~~INTRODUCTION}
%%%%%%%%%%%%%%%%%%%%%%%%%%%%%%%%%%%%%%%%%%%%%%%%%%%%
%
Paramagnetic-metal/ferromagnetic-insulator heterostructures provide a unique platform to explore spin-current phenomena such as spin pumping \cite{Kajiwara2010nature,magnon-spintronics}, spin-transfer torque \cite{Kajiwara2010nature,magnon-spintronics,SMR_Nakayama2013RPL,SMR_Hahn2013RPB,SMR_Althammer2013RPB,SMR_Vlietstra2013APL,Hamadeh2014PRL,Cornelissen2015NatPhys}, and spin Seebeck effect \cite{SSE_insulaor,SSE_Kikkawa2013PRL,spincaloritronics1,spincaloritronics2,SSE_review}. In this structure, itinerant electrons in the paramagnet and magnons in the ferromagnet interact with each other via the interfacial spin-exchange interaction \cite{Tserkovnyak05}. One of the most widely-used heterostructures for studying spin-current phenomena is a Pt/Y$_3$Fe$_5$O$_{12}$ (YIG) junction system. Pt is a paramagnetic metal exhibiting high spin-charge conversion efficiency due to its strong spin-orbit interaction \cite{ISHE_Azevedo,ISHE_Saitoh,ISHE_Costache,ISHE_Hoffmann,ISHE_Sinova}. YIG is a ferrimagnetic insulator with a high Curie temperature ($\sim 560~\textrm{K}$) \cite{YIG_Gilleo-Geller,SSE_Uchida2014PRX} and extremely high resistivity \cite{Metselaar}. Therefore, this structure enables pure detection of spin-current phenomena free from conduction-electrons' contribution in YIG. \par
In Pt/YIG junction systems, an unconventional magnetic field $H$ dependence of Hall effects has been observed \cite{Hunag2012PRL,Shimizu2013PRL,MR_Miao2014PRL,MR_Shiomi2014APL,Hall_Meyer2015APL,MR_Miao2016}. Although Pt films on non-magnetic substrates such as SiO$_2$ and Y$_3$Al$_5$O$_{12}$ (YAG) show a conventional $H$-linear response due to the normal Hall effect, the measured Hall effect in Pt/YIG systems exhibits two anomalous behaviors: (i) in a low-$H$ range ($H \lesssim 2~\textrm{kOe}$), the $H$ dependence of the Hall voltage looks reflecting the magnetization process of YIG and (ii) in a high-$H$ range much greater than the saturation field of YIG  ($\sim 2~\textrm{kOe}$), the magnitude of the Hall voltage nonlinearly increases with increasing $H$. The behavior (ii) becomes prominent at low temperatures especially below $100~\textrm{K}$. Since both the behaviors (i) and (ii) become more outstanding with decreasing the Pt thickness, these unconventional Hall effects should be related to interface effects at Pt/YIG.\par
To explain the unconventional Hall effects, several scenarios have been proposed. As a possible origin of the behavior (i), Chen {\it et al.} \cite{SMR_Chen2013RPB} and Zhang {\it et al.} \cite{AHE_Zhang2016PRL} independently suggested a role of nonequilibrium spin currents combined with interfacial spin-mixing effects. On the other hand, Huang {\it et al.} \cite{Hunag2012PRL,SSE_Qu2013PRL,XMCD_Lu2013PRL} and Guo {\it et al.} \cite{AHE_Guo2014PRB} pointed out a role of static Pt ferromagnetism induced by magnetic proximity effects at the Pt/YIG interface. The latter scenario is based on the fact that Pt is near the Stoner ferromagnetic instability \cite{Ibach,Mattheiss1980PRB,Tamura1995PRL}, and thereby it might be magnetized due to static proximity effects near the interface, as reported in various Pt/ferromagnetic-metal junction systems \cite{Vlachos2014,Suzuki2005PRB,Figueroa2014PRB,Kuschel2015PRL}. A possible mechanism for the behavior (ii) was discussed by Shimizu {\it et al.} \cite{Shimizu2013PRL} and Lin {\it et al.} \cite{MR_Lin2013APL} in terms of independent paramagnetic moments in Pt, which cause skew scatterings for itinerant electrons in Pt \cite{MR_Lin2013APL,Nagaosa2010RMP,Fert1972PRL,Fert1980text,Hamzic1980JMMM,Fert1981JMMM,Shinde2004PRL}. However, there is no direct evidence for the existence of such paramagnetic moments in Pt on YIG. \par
To study magnetic properties of Pt films on YIG, X-ray magnetic circular dichroism (XMCD) \cite{Stohr-text,Lovesey-text,Wende_review,Nakamura-Suzuki,Rogalev-Wilhelm_review} measurements have been conducted at Pt $L_3$- and $L_2$-edges. Gepr\"ags {\it et al.} \cite{XMCD_Geprags2012APL,XMCD_Geprags2013arXiv} tried to detect magnetic proximity effects at Pt/YIG interfaces by means of XMCD measurements, but did not observe any XMCD signals in Pt($1.6-10~\textrm{nm}$)/YIG-film samples at room temperature within the margin of experimental error ($<0.003 \pm 0.001~ \mu _{\rm B}$). In contrast, Lu {\it et al.} \cite{XMCD_Lu2013PRL} reported a finite XMCD signal in a Pt($1.5~\textrm{nm}$)/YIG-film sample at 300 K and 20 K, at which total magnetic moment values per Pt atom were estimated to be $0.054~ \mu _{\rm B}$ and $0.076~ \mu _{\rm B}$, respectively. These results indicate that magnetic properties of Pt on YIG are sensitive to the qualities of the Pt film and Pt/YIG interface. However, since all the previous studies focused only on the behavior (i), the XMCD measurements were performed only in the low field range ($H<6~\textrm{kOe}$) \cite{XMCD_Lu2013PRL,XMCD_Geprags2012APL,XMCD_Geprags2013arXiv}. 
In the present study, we measured XMCD in a Pt film on a YIG substrate at high magnetic fields and low temperatures. We observed a clear XMCD signal at the Pt $L_{3}$- and $L_{2}$-edges and found that the signal suggests induced paramagnetic moments in Pt, providing an important clue to understand the unconventional Hall effect in the Pt/YIG systems. \par
%
%
%
%%%%%%%%%%%%%%%%%%%%%%%%%%%%%%%%%%%%%%%%%%%%%%%%%%%%
\section{II.~~SAMPLE PREPARATION, CHARACTERIZATION, AND EXPERIMENTAL SETUP} \label{sec:procedure}
\begin{figure}[htb]
\begin{center}
\includegraphics{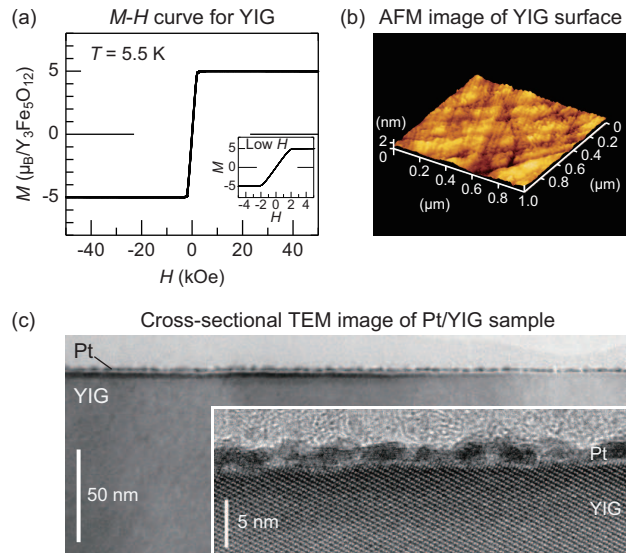}
\caption{(a) $M$-$H$ curve for the YIG slab when the magnetic field $H$ was applied along the [111] direction. The inset to (a) shows the magnified view of the $M$-$H$ curve in the low-$H$ range ($|H|<5~\textrm{kOe}$). The $M$-$H$ curve was measured with a vibrating sample magnetometer. (b) An AFM image of the polished (111) surface of the YIG slab before the Piranha treatment, where the surface roughness is $R_{\rm a} \sim 0.2~\textrm{nm}$. Similar AFM image and roughness value were confirmed also for the YIG surface with the Piranha treatment. (c) A cross-sectional TEM image of the Pt/YIG sample. The inset shows a blowup of the Pt/YIG interface.  }\label{fig:1}
\end{center}
\end{figure}

%%%%%%%%%%%%%%%%%%%%%%%%%%%%%%%%%%%%%%%%%%%%%%%%%%%%
%
To perform XMCD measurements, we prepared a junction system comprising a Pt film and a YIG substrate. The nominal thickness of the Pt film is $0.5~\textrm{nm}$, thinner than that used for the previous XMCD studies \cite{XMCD_Lu2013PRL,XMCD_Geprags2012APL,XMCD_Geprags2013arXiv}, because the unconventional Hall effect is enhanced with decreasing the Pt thickness \cite{XMCD_Lu2013PRL}. We used a single-crystalline YIG substrate grown by a flux method, commercially available from Ferrisphere Inc., USA, and cut it into a rectangular shape with the size of $8\times8\times1~\textrm{mm}^3$. As shown in Fig. \ref{fig:1}(a), the magnitude of the magnetization $M$ for the YIG at the temperature $T=5.5~\textrm{K}$ increases with increasing the magnetic field $H$ from zero and saturates at around $H = 2~\textrm{kOe}$, where the saturation $M$ value was observed to be $\sim $ 5$~\mu_{\rm B}$, consistent with a literature value \cite{YIG_Gilleo-Geller,SSE_review}. Before putting a Pt film, the $8\times8~\textrm{mm}^2$ surface [(111) surface] of the YIG slab was mechanically polished with sand papers and alumina slurry; the resultant surface roughness of the YIG slab is $R_{\rm a} \sim 0.2~\textrm{nm}$, as shown in an atomic force microscope (AFM) image in Fig. \ref{fig:1}(b). The YIG slab was cleaned with acetone and methanol in an ultrasonic bath and, then, cleaned with so-called Piranha etch solution (a mixture of H$_2$SO$_4$ and H$_2$O$_2$ at a ratio of 1:1) \cite{MR_Shiomi2014APL,Jungfleisch2013APL}. 
The Piranha etch solution may remove organic matter attached to the YIG surface \cite{Jungfleisch2013APL} and hence does not affect the surface roughness and morphology of the YIG, which were confirmed by our AFM measurements. 
We deposited a Pt film on the YIG surface by rf magnetron sputtering with an rf power density of $0.88~\textrm{W/cm}^2$ in an Ar atmosphere of $4.8~\textrm{mTorr}$ at room temperature, which results in a deposition rate of $0.028~\textrm{nm/sec}$. 
The Pt/YIG interface and the morphology of the Pt film were evaluated by means of transmission electron microscopy (TEM). As shown in Fig. \ref{fig:1}(c), we confirmed a flat YIG surface and also a clear Pt/YIG interface. Due to its island-growth nature  \cite{Figueroa2014PRB}, the Pt film was found to consist of discontinuous clusters with a typical thickness of $\sim 2~\textrm{nm}$, which may reduce an interface-to-volume ratio compared to ideal flat films and may make proximity-induced Pt magnetic moment densities be cluster-size dependent. Nevertheless, the observed clear contact between the Pt and YIG allows us to evaluate the characteristic of the magnetic proximity effect in the present system at high magnetic fields and low temperatures. We also prepared a control sample comprising a (nominally) 0.5-nm-thick Pt film on a non-magnetic YAG (111) substrate with the size of $8\times8\times0.5~\textrm{mm}^3$. The Pt films on the YIG and YAG substrates were deposited at the same time in our sputtering chamber. \par
%The transmission electron microscopy (TEM) 
%In Figs. \ref{fig:1}(c) and \ref{fig:1}(d)
%\cite{Figueroa2014PRB}
% the Pt was found to form small clusters  
%  which could be a source  reduce the contact region beween the Pt and YIG
%
%
\begin{figure}[htb]
\begin{center}
\includegraphics{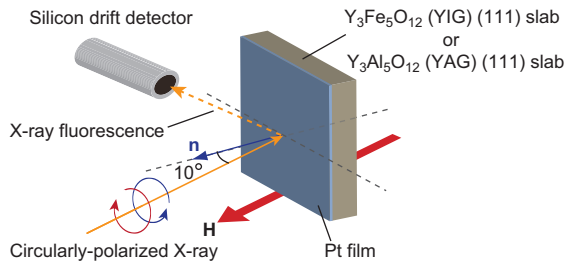}
\caption{A schematic illustration of the experimental setup for the XMCD measurements. A magnetic field, ${\bf H}$, parallel or antiparallel to the incoming X-ray beam, was applied to the Pt/YIG or Pt/YAG sample at an angle of $10^{\circ}$ to the normal direction ${\bf n}$ of the sample surface.}\label{fig:2}
\end{center}
\end{figure}
Figure \ref{fig:2} shows the experimental setup for the XMCD measurements. The XMCD experiments were performed at the beam line BL39XU of SPring-8 synchrotron radiation facility using the fluorescence detection mode \cite{Suzuki2005PRB}. X-ray absorption spectra (XAS) were recorded at the Pt $L_3$-edge ($2p_{3/2}\rightarrow 5d$ valence, $11570~\textrm{eV}$) and Pt $L_2$-edge ($2p_{1/2}\rightarrow 5d$ valence, $13282~\textrm{eV}$) with circularly-polarized X-rays while reversing their helicity at $1~\textrm{Hz}$. Here, by measuring the XAS for the two circular polarizations almost at the same time, we can exclude time-dependent artifacts. The circularly-polarized X-rays with a high degree of polarization ($\geqq 95~\% $) were generated by using a diamond X-ray phase retarder. The Pt $L_{\alpha}$ (for the Pt $L_3$-edge) and $L_{\beta}$ (for the Pt $L_2$-edge) fluorescences were measured with a silicon drift detector. In order to perform XMCD measurements in the configuration similar to the set-up for the Hall measurements, an out-of-plane magnetic field, ${\bf H}$, was applied to the samples using a split-type superconducting magnet, where the angle between ${\bf H}$ and the direction normal to the sample surface ${\bf n}$ was set to be $10^\circ$. The direction of the angular momentum vector of the incident X-ray beam was parallel or antiparallel to the ${\bf H}$ direction. The XMCD signal was obtained by taking a difference of the XAS recorded with the opposite helicities at $H= 50~\textrm{kOe}$. The measurements of element-specific magnetization curves were performed at the constant energy of $E = 11568~\textrm{eV}$ (the XMCD peak position for the Pt $L_{3}$-edge) with sweeping the $H$ value between $\pm ~50~\textrm{kOe}$. The temperature was fixed at $T=5.5~\textrm{K}$ in all the measurements. \par
%
%%%%%%%%%%%%%%%%%%%%%%%%%%%%%%%%%%%%%%%%%%%%%%%%%%%%
\section{III.~~RESULTS AND DISCUSSION}
%%%%%%%%%%%%%%%%%%%%%%%%%%%%%%%%%%%%%%%%%%%%%%%%%%%%
%
\begin{figure}[htb]
\begin{center}
\includegraphics{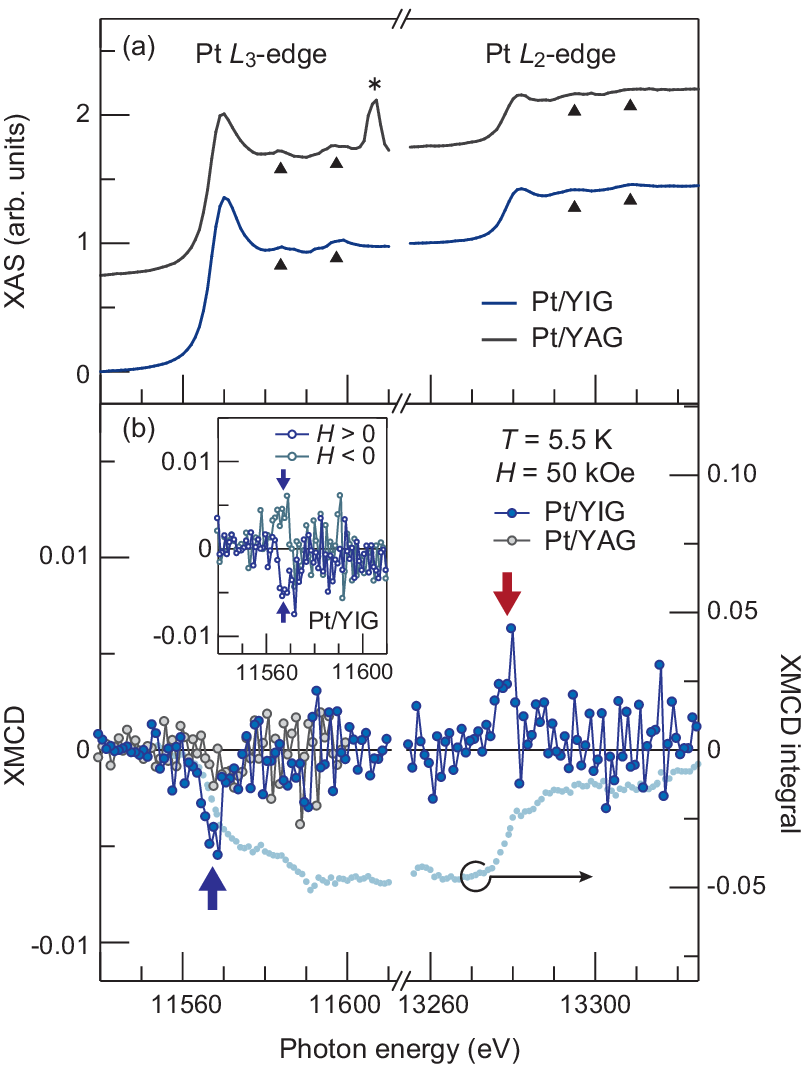}
\caption{(a) The normalized XAS for the Pt $L_{3}$- and $L_{2}$-edges of the Pt/YIG and Pt/YAG samples at $T = 5.5~\textrm{K}$ and $H= 50~\textrm{kOe}$. The XAS edge jump, defined as the difference in the XAS intensity between $11540~\textrm{eV}$ ($13255~\textrm{eV}$) and $11610~\textrm{eV}$ ($13325~\textrm{eV}$), is normalized to 1 ($2.22^{-1}$) for the $L_3$-edge ($L_2$-edge) according to Refs. \onlinecite{Mattheiss1980PRB} and \onlinecite{Bartolome2009PRB}. The XAS offset value for the $L_3$-edge ($L_2$-edge) of the Pt/YIG sample is set to be 0 (1) \cite{Figueroa2014PRB,XMCD_Geprags2012APL,XMCD_Geprags2013arXiv,Bartolome2009PRB,Rogalev-Wilhelm_review}. Similarly, the XAS offset value for the $L_3$-edge ($L_2$-edge) of the Pt/YAG sample is set to be 0 (1) and then, for clarity, is shifted along the vertical axis by a constant value of 0.75 (0.75). The triangles indicate the oscillation structures in the EXAFS region. The peak structure marked with an asterisk in the XAS at $11607~\textrm{eV}$ for the Pt/YAG sample appears due to elastic diffraction from the YAG substrate \cite{XMCD_Geprags2013arXiv}. (b) The XMCD spectra and their integral for the Pt $L_3$- and $L_2$-edges of the Pt/YIG sample. The blue (red) arrow indicates the negative (positive) XMCD signal at the Pt $L_3$-edge ($L_2$-edge). The inset to (b) shows the XMCD spectra for the Pt $L_3$-edge of the Pt/YIG sample measured at $H= \pm ~50~\textrm{kOe}$, where the blue arrows indicate the XMCD signal. In (b), the XMCD spectra for the Pt $L_3$-edge of the Pt/YAG sample are also plotted.}\label{fig:3}
\end{center}
\end{figure}
Figure \ref{fig:3}(a) shows the XAS for the Pt $L_3$- and $L_2$-edges of the Pt/YIG and Pt/YAG samples, where the XAS edge jump is normalized to 1 ($2.22^{-1}$) for the $L_3$-edge ($L_2$-edge) \cite{Mattheiss1980PRB,Bartolome2009PRB}. The whiteline intensity, the ratio of the absorption maximum at the $L_3$-edge to the edge jump, is estimated to be 1.35 for the Pt/YIG sample. This value is consistent with that for the Pt($1.6-10~\textrm{nm}$)/YIG systems reported in Refs. \onlinecite{XMCD_Geprags2012APL} and \onlinecite{XMCD_Geprags2013arXiv}, indicating a mainly metallic state of our Pt films (note that the whiteline intensity was reported to be $1.25-1.30$, $1.50$, and $2.20$ for metallic Pt foil, PtO$_{1.36}$, and PtO$_{1.6}$, respectively \cite{XMCD_Geprags2013arXiv}). The quality of our Pt films was also confirmed by the clear oscillation in the extended X-ray absorption fine structure (EXAFS), marked with triangles in Fig. \ref{fig:3}(a), which is identical with metallic Pt \cite{XMCD_Geprags2013arXiv}.\par
Figure \ref{fig:3}(b) shows the XMCD spectra for the Pt/YIG sample at $T=5.5~\textrm{K}$ and $H=50~\textrm{kOe}$. We observed small but finite XMCD signals with a negative sign at the $L_3$-edge ($11568~\textrm{eV}$) and with a positive sign at the $L_2$-edge ($13280~\textrm{eV}$). The sign of the XMCD signal was found to be reversed by reversing the ${\bf H}$ direction [see the inset to Fig. \ref{fig:3}(b)]. These results confirm that the observed XMCD signals are of magnetic origin; the Pt film on YIG is slightly magnetized under such a high field and the ${\bf M}$ direction of the Pt responds to the ${\bf H}$ direction. \par
Importantly, the XMCD intensity relative to the XAS edge jump at $H= 50~\textrm{kOe}$ for the Pt/YIG sample ($\sim 5\times10^{-3}$) is several times greater than that for the Pt/YAG sample ($\sim 1\times10^{-3}$) [Fig. \ref{fig:3}(b)] and a Pt foil ($\sim 1\times10^{-3}$) \cite{Bartolome2009PRB}. The result indicates that, due to the YIG contact, the Pt film acquires magnetic moments greater than the Pauli paramagnetic moments \cite{Rogalev-Wilhelm_review,Bartolome2009PRB,Bartolome2013}. 
By XMCD sum-rule analysis \cite{sum-rule1,sum-rule2}, the averaged Pt magnetic moment per Pt atom in the whole volume of the Pt film on YIG at $H=50~\textrm{kOe}$ was estimated to be $m_{\rm tot} = m_{\rm orb} +m_{\rm spin}^{\rm eff}  = 0.0212 \pm 0.0015 ~\mu_{\rm B}$, where $m_{\rm orb} = 0.0017 \pm 0.0006 ~\mu_{\rm B}$ and $m_{\rm spin}^{\rm eff} = 0.0195 \pm 0.0014 ~\mu_{\rm B}$ are the orbital and effective spin magnetic moments per Pt atom, respectively (see Appendix). \par
\begin{figure}[htb]
\begin{center}
\includegraphics{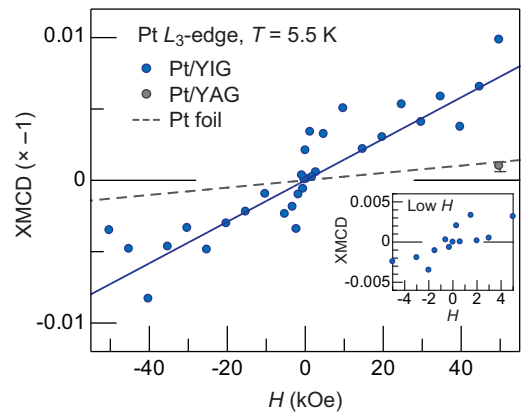}
\caption{The ESM curve at the Pt $L_{3}$-edge of the Pt/YIG sample at $T = 5.5~\textrm{K}$ and the fixed photon energy of $11568~\textrm{eV}$, where the blue dots and blue solid line represent the experimental data and the linear-fitting result, respectively. The XMCD data at the Pt $L_{3}$-edge of the Pt/YAG sample (gray circle), estimated from Fig. \ref{fig:3}(b), and of a Pt foil at $T = 10~\textrm{K}$ (gray dashed line), taken from Ref. \onlinecite{Bartolome2009PRB}, are also plotted. The inset shows the magnified view of the ESM curve of the Pt/YIG sample in the low-$H$ range ($|H|<5~\textrm{kOe}$). The XMCD values of the vertical axes are multiplied by $-1$ for clarity, since the sign of the XMCD signal at the Pt $L_{3}$-edge is negative [see Fig. \ref{fig:3}(b)]}\label{fig:4}
\end{center}
\end{figure}

 
%.
To clarify the $H$ dependence of the magnetic moments in the Pt film on YIG, we measured the element-specific magnetization (ESM) curve at the Pt $L_3$-edge for the same Pt/YIG sample at $T=5.5~\textrm{K}$ (see Fig. \ref{fig:4}). Importantly, the magnitude of the XMCD signal was observed to increase gradually and almost linearly with $H$. This behavior cannot be explained by the Pt ferromagnetism induced by the magnetic proximity effect, since the XMCD signal due to the proximity-induced Pt ferromagnetism, if it exists, reflects the $M$-$H$ curve of the YIG substrate surface and the $M$ of the YIG saturates at around $H=2~\textrm{kOe}$ [see Fig. \ref{fig:1}(a)] \cite{comment1}. 
Significantly, the slope of the XMCD signal with respect to $H$ for the Pt/YIG sample was found to be $5.7$ times greater than that for the Pt foil where only the Pauli paramagnetism of Pt appears (see Fig. \ref{fig:4}). The result shows that the large XMCD slope observed in the Pt/YIG sample cannot be explained also by the simple Pauli paramagnetic moments. \par
The above ESM results suggest that the Pt film on YIG acquires paramagnetic moments greater than the Pauli paramagnetic moments under such high magnetic fields. In the following, we discuss a possible origin of the induced moments. 
In general, induced Pt magnetic moments in a Pt/ferromagnet interface are attributed to direct interaction between $5d$-electrons in the Pt and spin-polarized electrons in the ferromagnet \cite{Wende_review}, and thus the $M$-$H$ (ESM) curve for the induced magnetic moment in the Pt reflects the magnetization process of the magnet which is coupled to the Pt \cite{Figueroa2014PRB}. This indicates that the large paramagnetic moments observed in the Pt/YIG sample may originate from magnetic coupling between Pt and interfacial magnetic moments at the Pt/YIG interface whose magnetization process exhibits the paramagnetic behavior. 
The followings are possible candidates which may explain the appearance of such interfacial magnetic moments: (1) interdiffusion of ions and alloying at the Pt/YIG interface; (2) local amorphous structures of the YIG surface, which were proposed to exhibit magnetic properties different from the YIG bulk \cite{Saiga2014APEX,Song2015APL}; (3) Y, Fe, and O vacancies of the YIG surface \cite{Liang2016DFT,Song2015APL,Lomako2011CrystallogrRep}; and  (4) unintentional formations of free Fe ions on the YIG surface \cite{MR_Lin2013APL}. Future detailed studies on magnetic properties of the Pt/YIG interface are desirable for full understanding of the origin of the observed paramagnetic behavior in Pt. 
Furthermore, the relevance between the Pt paramagnetic moments and the cluster-like-growth film structure [Fig. \ref{fig:1}(c)] and possible cluster-size dependence of the Pt paramagnetic moment densities are issues to be addressed by experimental and theoretical approaches.
\par
Finally, we comment on the relevance between the induced paramagnetic moments and the unconventional Hall effect under high magnetic fields observed in Pt/YIG systems. 
We found that the gradual increasing behavior of the induced Pt paramagnetic moments with $H$ shown in Fig. \ref{fig:4} is similar to that of the Hall effect in the Pt/YIG systems at high magnetic fields and low temperatures \cite{MR_Miao2014PRL,MR_Shiomi2014APL,MR_Miao2016}, suggesting that the paramagnetic moments are relevant to the Hall effect.  
It has been reported that, if paramagnetic moments exist in a conductor, they induce anomalous Hall effects due to the skew-scattering mechanism \cite{Nagaosa2010RMP,Fert1972PRL,Fert1980text,Hamzic1980JMMM,Fert1981JMMM,Shinde2004PRL}. 
This suggests that the unconventional Hall effect under high magnetic fields in the Pt/YIG systems can be attributed to the induced paramagnetic moments. 
This scenario is supported also by the recent reports by Miao {\it et al.} \cite{MR_Miao2014PRL,MR_Miao2016}. They reported that the Hall effect in the Pt/YIG systems under high magnetic fields is well consistent with that in a diluted-Fe-doped Pt film and a pure Pt film on a diluted-Fe-doped SiO$_2$ film \cite{MR_Miao2014PRL,MR_Miao2016}, where Fe impurities exhibit a paramagnetic behavior in Pt \cite{Craig1962PhysRev,Kitchens1965PhysRev,Maley1967JAP,Graham1968JAP,Nieuwenhuys1974,Stepanyuk1996PRB,Herrmannsdorfer1996JLowTempPhys}, suggesting a role of the induced paramagnetic moments in the mechanism of the unconventional Hall effects in the Pt/YIG systems. \par
%
%
%% 
%%%%%%%%%%%%%%%%%%%%%%%%%%%%%%%%%%%%%%%%%%%%%%%%%%%%
\section{IV.~~CONCLUSION}
%%%%%%%%%%%%%%%%%%%%%%%%%%%%%%%%%%%%%%%%%%%%%%%%%%%%
%
In this study, we estimated the magnetic moments in an ultra-thin Pt film on a ferrimagnetic insulator Y$_3$Fe$_5$O$_{12}$ (YIG) at high magnetic fields (up to $50~\textrm{kOe}$) and at low temperatures ($5.5~\textrm{K}$) by means of X-ray magnetic circular dichroism (XMCD). We observed an XMCD signal due to magnetic moments in the Pt film at the Pt $L_{3}$- and $L_{2}$-edges. The measurements of element-specific magnetization curves at the Pt $L_{3}$-edge reveal unconventional paramagnetic moments of which the intensity is greater than that expected from the Pauli paramagnetism of Pt. Our experimental results reported here provide an important clue in unraveling the nature of the unconventional Hall effects observed in Pt/YIG systems at high magnetic fields and low temperatures. \par
%
%%%%%%%%%%%%%%%%%%%%%%%%%%%%%%%%%%%%%%%%%%%%%%%%%%%%
\section*{ACKNOWLEDGMENTS}
%%%%%%%%%%%%%%%%%%%%%%%%%%%%%%%%%%%%%%%%%%%%%%%%%%%%
%
%
The synchrotron radiation experiments were performed at the beam line BL39XU of SPring-8 synchrotron radiation facility with the approval of the Japan Synchrotron Radiation Research Institute (JASRI) (Proposal Nos. 2013B1910, 2014A1204, 2015A1178, and 2015A1457). The authors thank K. S. Takahashi, S. Shimizu, Y. Shiomi, T. Niizeki, T. Ohtani, T. Seki, S. Ito, D. Meier, T. Kuschel, and S. T. B. Goennenwein for valuable discussions and N. Kawamura for experimental assistance. This work was supported by ERATO ``Spin Quantum Rectification Project'' (No. JPMJER1402) and PRESTO ``Phase Interfaces for Highly Efficient Energy Utilization'' (No. JPMJPR12C1) from JST, Japan, Grant-in-Aid for Scientific Research on Innovative Area ``Nano Spin Conversion Science'' (No. JP26103005), Grant-in-Aid for Scientific Research (A) (No. JP15H02012) from JSPS KAKENHI, Japan, NEC Corporation, and The Noguchi Institute. T.K. is supported by JSPS through a research fellowship for young scientists (No. JP15J08026). \par
%
%
%%%%%%%%%%%%%%%%%%%%%%%%%%%%%%%%%%%%%%%%%%%%%%%%%%%%
\section*{APPENDIX: XMCD SUM RULE ANALYSIS}
%%%%%%%%%%%%%%%%%%%%%%%%%%%%%%%%%%%%%%%%%%%%%%%%%%%%
We estimate the average orbital, effective spin, and total magnetic moments, $m_{\rm orb}$,  $m_{\rm spin}^{\rm eff}$, and $m_{\rm tot}$, per Pt atom in the whole volume of the Pt film on YIG from the integrated XAS and XMCD spectra by using the following sum rules \cite{sum-rule1,sum-rule2}:
%[width=7cm]
%
\begin{figure}[htb]
\begin{center}
\includegraphics{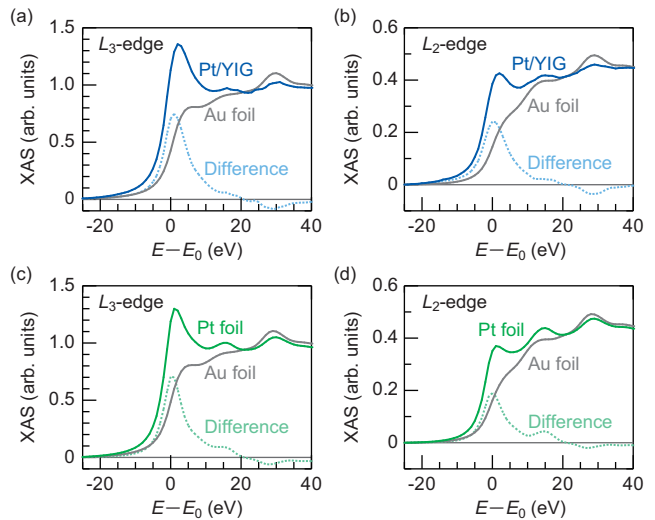}
\caption{[(a) and (b)] The normalized XAS for the (a) Pt $L_{3}$-edge and (b) Pt $L_{2}$-edge of the Pt/YIG sample. [(c) and (d)] The normalized XAS for the (c) Pt $L_{3}$-edge and (d) Pt $L_{2}$-edge of the Pt foil. The horizontal axes are shifted with respect to the energy at the inflection point $E_0$ for the Pt spectra. In (a) and (b) [(c) and (d)], the XAS for the Au $L_{3}$- and $L_{2}$-edges of the Au foil and difference in the XAS between the Pt/YIG sample (Pt foil) and Au foil are also plotted, where the energy scale of the Au XAS is expanded by a factor of 1.07 to take into account the difference in the lattice constant between Pt and Au \cite{Grange1998PRB} and is shifted in energy on the basis of the oscillation structures in the EXAFS region of the Pt spectra.}\label{fig:5}
\end{center}
\end{figure}
%[width=8cm]
% is shifted in energy to coincide with the EXAFS oscillations between the XAS of the Au foil and Pt film.
%
\begin{equation}\label{equ:sum-rule1}
m_{\rm orb} = -\frac{2}{3} \frac{\Delta I_{L_{3}} + \Delta I_{L_{2}}}{I_{L_{3}}+I_{L_{2}}}n_{\rm h} \mu _{\rm B},
\end{equation}
\begin{equation}\label{equ:sum-rule2}
m_{\rm spin}^{\rm eff} = -\frac{\Delta I_{L_{3}}-2\Delta I_{L_{2}}}{I_{L_{3}}+I_{L_{2}}}n_{\rm h} \mu _{\rm B},
\end{equation}
where $I_{L_{3}}$ ($I_{L_{2}}$) is the XAS integral summed over the Pt $L_{3}$-edge ($L_{2}$-edge) after subtracting the contribution coming from electron transitions to the continuum, $\Delta I_{L_{3}}$ ($\Delta I_{L_{2}}$) is the integral of the XMCD spectra for the Pt $L_{3}$-edge ($L_{2}$-edge), $n_{\rm h}$ is the number of holes in the Pt $5d$ band, and $\mu_{\rm B}$ is the Bohr magneton. 
We first calculate the value of the XAS integral $r=I_{L_{3}}+I_{L_{2}}$ for the Pt/YIG sample. To remove the X-ray absorption due to the electron transitions to the continuum, we subtract the XAS of an Au foil from those of the Pt film according to the method proposed in Refs. \onlinecite{Rogalev-Wilhelm_review}, \onlinecite{Bartolome2009PRB}, and \onlinecite{Grange1998PRB}. To do this, the energy scale of the XAS of the Au foil was expanded by a factor of 1.07 to take into account the difference in the lattice constant between Pt and Au and to align in energy with that of the Pt film. The XAS of the Au foil is normalized to the edge jump (see Fig. \ref{fig:5}). The area differences between the Pt-film and Au-foil spectra for both the $L_{3}$- and $L_{2}$-edges, i.e., the blue dashed curves in Figs. \ref{fig:5}(a) and \ref{fig:5}(b), were integrated between $-20~\textrm{eV} \leqq E-E_0 \leqq 20~\textrm{eV}$, where $E_0$ is the energy at the inflection point of the Pt XAS. As a result, the XAS integral for the Pt/YIG sample was estimated to be $r=10.2~\textrm{eV}$.
The absolute value of $n_{\rm h}$ for the Pt film on YIG  was calculated following the methodology described in Ref. \onlinecite{Bartolome2009PRB}. First, we determined the scaling factor $C^{-1}={\tilde n}_{\rm h}/r_{\rm {Pt-foil}}=0.112~\textrm{holes/eV}$, where ${\tilde n}_{\rm h}=n_{\rm h}^{\rm Pt} - n_{\rm h}^{\rm Au} = 0.98$ is the difference between the $5d$ holes of the Pt metal $n_{\rm h}^{\rm Pt}(=1.73)$ and the Au metal $n_{\rm h}^{\rm Au}(=0.75)$. $r_{\rm {Pt-foil}} = 8.72~\textrm{eV}$ is the XAS-integral value estimated from the XAS of the Pt foil [Figs. \ref{fig:5}(c) and \ref{fig:5}(d)]. Second, we calculated the $n_{\rm h}$ value of the Pt film as $n_{\rm h} = n_{\rm h}^{\rm Au} + C^{-1}r = 1.89$. 
Using the estimated $r=I_{L_{3}}+I_{L_{2}}$, $n_{\rm h}$, and the integrals of the XMCD spectra for the Pt $L_{3}$- and $L_{2}$-edges ($\Delta I_{L_{3}}$ and $\Delta I_{L_{2}}$) [see Fig. \ref{fig:3}(b)], the orbital and effective spin magnetic moments were respectively determined to be $m_{\rm orb} = 0.0017 \pm 0.0006 ~\mu_{\rm B}$ and $m_{\rm spin}^{\rm eff} = 0.0195 \pm 0.0014 ~\mu_{\rm B}$ from Eqs. (\ref{equ:sum-rule1}) and (\ref{equ:sum-rule2}). This result leads to the total magnetic moment of $m_{\rm tot} = m_{\rm orb} +m_{\rm spin}^{\rm eff}  = 0.0212 \pm 0.0015 ~\mu_{\rm B}$. 
We note that the derived magnetic moments should be regarded as the values averaged over the whole volume of the Pt layer, since the XMCD gives a mean polarization of the Pt film. \par
%
%

%\newpage
%
%
%
%
%
\end{document}